\documentclass{amsart}

\usepackage{epsf}
\usepackage{graphicx,graphics,subfigure}

\pdfoutput=1

\begin{document}

%


\title{Robust Soldier Crab Ball Gate}

\author[Yukio-Pegio Gunji]{Yukio-Pegio Gunji}
\address{Department of Earth \& Planetary Sciences, Kobe University, Kobe 657-8501, Japan}
\author[Yuta Nishiyama]{Yuta Nishiyama} 
\address{Center for Complex Systems, Waseda University, Shinjuku, Tokyo 169-8050, Japan}
\author[Andrew Adamatzky]{Andrew Adamatzky}
\address{Unconventional Computing Centre, UWE, Bristol, United Kingdom}

\maketitle 

   

\begin{abstract} 
Soldier crabs \emph{Mictyris guinotae} exhibit pronounced swarming behaviour. The swarms of the crabs tolerant of perturbations. In computer models and laboratory experiments we demonstrate that swarms of soldier crabs can implement logical gates when placed in a geometrically constrained environment.
\end{abstract}

\section{Introduction} 
\label{intro}

All natural processes can be interpreted in terms of computations. To implement a logical gate in a chemical, physical or biological spatially extended media one must assign Boolean variables to disturbances, defects or localizations travelling in the media, collide these travelling patterns and convert outcomes of their collision into resultant logical operations. This is how collision-based computers work~\cite{adamatzky_2007, adamatzky_2010}. Now classical examples of experimental laboratory unconventional computing include the Belousov-Zhabotinsky (BZ) medium and the wslime 
mould of \emph{Physarum polycephalum}. In BZ excitable medium logical variables are represented by excitation waves, which interact with each other in the geometrically constrained substrate   or  'free-space' 
substrate~\cite{adamatzky_2005,yoshikawa_2009,toth_1994,toth_1995}. Slime mould is capable of solving many computational problems including maze and adaptive networks~\cite{nakagaki_2000, adamatzky_2011}. In the case of ballistic computation \cite{margolus_1984} slime moulds implement collision computation when two slimes are united or avoided with each other dependent on the gradient of attractor and inhibitor. We previously suggested that the slime mould logical gate is robust against external perturbation \cite{tsuda_2004,adamatzky_2010}. To expand the family of unconventional spatially extended computers we studied swarming behaviour of soldier 
crabs~\emph{Mictyris guinotae} and found that compact propagating groups of crabs emerge and sustain under noisy external stimulation. We speculated that swarms can behave similarly to billiard balls and thus implement basic circuits of collision-based computing, results of our studies are presented below.

\section{Swarming of soldier crabs}

Soldier crabs \emph{Mictyris guinotae} inhabit flat lagoons and form huge colonies of several hundreds and sometimes hundreds of thousands of crabs.  In field expeditions to Funaura Bay, Iriomote Island, Japan, we observed 
a peculiar wandering behaviour of crabs. A front part of their swarm is driven by inherent turbulence. The arrangement of individuals is always  changing. A single crab or a small of group of crabs do not usually enter the water, however 
a large swarm enters the water and cross a lagoon without hesitation. The large swarm crossing the water consists of an active front and passive tail. The crabs in the tails simply follow the crabs at the front. We assumed there are two types of neighbourhoods: one for positive interaction and another for monitoring and following flock-mates.

\begin{figure}
\centering
\includegraphics[width=0.7\textwidth]{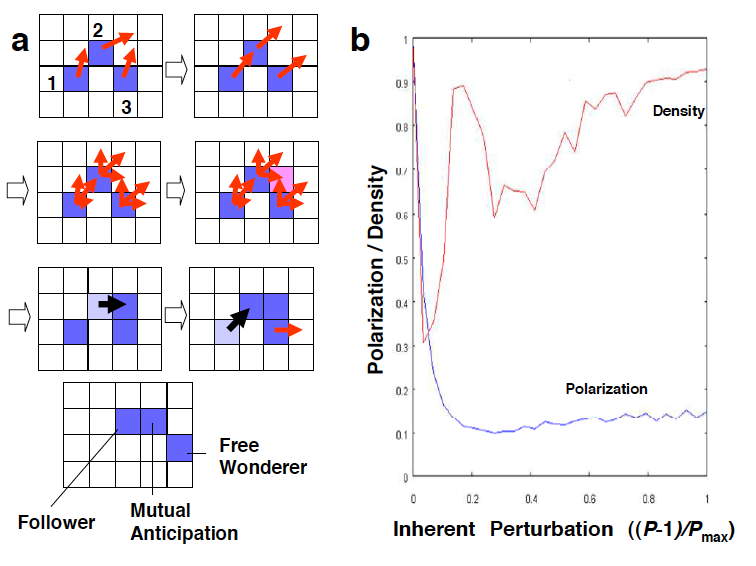}
\caption{ a. Schematic diagram of transitions in model swarm. Blue cells represent individuals. A pink cell represents the site with high popularity. Pale blue cells represent fresh absent cells. Each cell designates a site in a space. See text for details. b. Polarisation and density plotted against inherent perturbation. The inherent perturbation is defined by the number of potential transitions normalised by the maximum number of potential transition.}
\label{fig01}
\end{figure}

To implement these behaviours in a model, we introduce a number of potential transitions for each individual that can be employed to mutual anticipation \cite{rosen_1985}. Each individual has its own principal vector representing its own velocity, accompanied with $P$ numbers potential transitions in a range restricted by angle $\alpha$. The neighbourhood has dual roles: the extended body of an individual for active interaction and the local space for passive monitoring other individuals. 

Figure \ref{fig01} shows a typical transition of our model swarm of which individuals are numbered from 1 to 3. First velocity matching is applied to the principal vector by averaging all flock-mate velocities in the neighbourhood, NM that is here defined by Moore neighbourhood extended by next nearest neighbouring cells (Fig. 1a, left and centre in top row). If targets of some potential transitions are overlapped at a particular cell (Fig. 1a right in top row), the overlapping is counted as ÒpopularityÓ. In Fig. 1a, left in centre row, transitions from the individual 2 and 3 are overlapped at a pink cell. It means that the pink cell has popularity with 2. If some potential transitions reach high popular sites beyond the threshold of popularity (here that is defined by 1), an individual moves to the site with the highest popularity. In Fig. 1a, centre in centre row, the individual 2 moves to the highest popular cell. If several individuals intend to move to the same site, one individual is randomly chosen, and others move to the second best site. In Fig. 1a, both individual 2 and 3 can move to the popular site. The individual 2 was randomly chosen. This rule implements individualÕs mutual anticipation. Even for a human, he or she can avoid collision in a crowded pedestrian by anticipation \cite{gunji_2011}. We implement this kind of behaviour by the mutual anticipation. If there is no popular site in any targets of potential transitions and other individual in the neighbourhood of Following, NF that is here defined by Moore neighbourhood, moves due to the mutual anticipation, the individual move to occupy the absent cell generated by the mutual anticipation. Namely, it follows the predecessor (Fig. 1a, right in the centre row). If an individual is obeyed neither to the mutual anticipation nor to following, it moves in the direction of one of potential transitions randomly chosen. It is called free wanderer. In Fig. 1a, right in the centre row, the individual 3 is a free wanderer.

Fig. 1b shows how intrinsic turbulence is generated and maintained in our swarm model. Polarisation is usually used to estimate the order of coherence for a flock and a school. Polarisation is defined by the length of summation for the all agentsÕ velocities (principal vector in our model). Density is defined by the average number of agents located in the neighbourhood of each agent. In Fig.1ab polarisation and density are normalised by maximal polarisation and maximal density, respectively. The inherent perturbation is defined by the number of potential transitions minus 1, normalised by maximal number of potential transitions that is 30 in Fig. 1b. No external noise is coupled with the transition rule for an agent.

If $P=1$ (i.e., inherent perturbation is 0.0) mutual anticipation cannot be applied to each agent, and the model mimics the BOIDS. Once agents are aggregated by flock centring, velocity matching makes a flock with high polarisation, and a flock moves as a mass. Thus it reveals both high polarisation and high density. If $P=2$, each agent has two potential transitions. Although it is possible that potential transition can contribute to swarming, overlapping of targets of potential transitions is too difficult to achieve mutual anticipation. The effect of multiple transitions results only in random choice of potential transitions for each agent. Thus, both density and polarisation are very small. As the inherent perturbation increases, density increases and polarisation decreases. High density and low polarisation are maintained where inherent perturbation is larger than 0.5.

In BOIDS and SPP, highly dense flock is achieved only by high polarisation. The more highly polarised flock is, the denser flock is. By contrast, low polarisation breaks the coherence of a flock. The velocity matching is linearly coupled with external perturbation in BOIDS and SPP. If the polarisation is plotted against external perturbation, the polarisation decreases as the perturbation decreases. Because polarisation shows phase transition dependent on the external perturbation, the polarisation is regarded as an order parameter. Thus the density of a flock is also decreased as the external perturbation decreases. In this framework, the noise disturbs the coherence of a flock, which reveals not robust but stable flock. By contrast, in our model inherent perturbation positively contributes to generate of a coherent and dense swarm and can implement robust swarms, which can used analogous to billiard balls in  collision-based circuitry.

\section{Collision computing by crab swarm}

\subsection{Simulation model}

When a swarm of soldier crabs is set in a corridor, it is expected that a swarm acts as a robust ball and goes straight, and that two swarm balls are united into one ball after the collision. Because of velocity matching, a swarm ball resulting from fusion of two balls has a summary velocity of two colliding balls.

\begin{figure}
\centering
\includegraphics[width=0.7\textwidth]{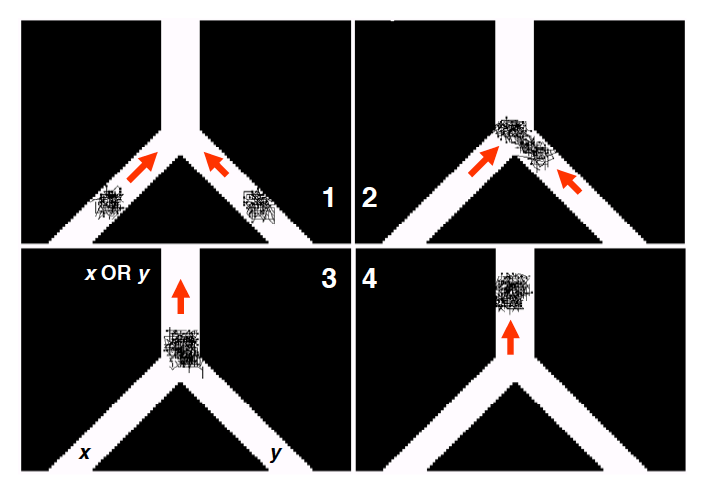}
\caption{ A series of snapshots (1, 2, 3 and 4) in {\sc OR} gate of swarm balls. A swarm ball located at $x$- and $y$-position consists of 40 agents, respectively. Each agent is represented by a square with its 5-steps-trajectories. Red arrows represent the direction of motion of a swarm ball.}
\label{fig02}
\end{figure}

Figure \ref{fig02} shows {\sc OR} gate of the collision computing implemented by swarm balls. In this simulation, black regions are walls which an agent cannot invade. Agents can move in a white area. An agent moves in following the rule mentioned before, where each agent has a tendency to move along the wall if it is close to the wall. This tendency is consistent with our observation for soldier crabs. In a corridor, they move along the wall. Because they have a tendency to move together, a swarm generated close to the wall of the corridor propagates along the wall. If a soldier crab is not close to the wall, it moves freely. In this {\sc OR} gate an agent is assumed to move upward along the wall in Fig. 2, if it is close to the wall.

When agents are set in the position of input, agents are aggregated into one swarm moving along the wall. In this gate,  a swarm initiated at $x$- or $y$-position moves upward along the central corridor after the swarm encounters the central corridor. It is easy to see that $(x, y) = (0, 0)$ leads to $x$ {\sc OR} $y=0$ and that $(x, y) = (0, 1)$ and 
$(1, 0)$ lead to $x$ {\sc OR}$ y =1$, where the presence of agents represents 1 and the absence represents 0. Fig. 2 shows a series of snap shots of the behaviour of {\sc OR} gate when two swarm balls are set at the $x$- and $y$-position. It shows that $(x, y) = (1, 1)$ leads to $x$ {\sc OR} $y =1$. 

We can estimate the effect of external perturbation if the agents' transition rule is coupled with external perturbation. As shown in Fig. 2, inherent noise can positively contribute to generate and maintains a robust swarm. Inherent noise in our model reveals multiple potential transitions, and one of them is always chosen in the transition of each agentÕs location. The inherent noise, therefore, cannot be distinguished from external noise. It means that even external noise can contribute to generate robust swarm balls while the direction of move of a swarm ball cannot be controlled. However unstable the direction of a swam ball is, the corridor of the {\sc OR} gate is one way and then the output 1 resulting from a swarm ball is robust.

\begin{figure}
\centering
\includegraphics[width=0.7\textwidth]{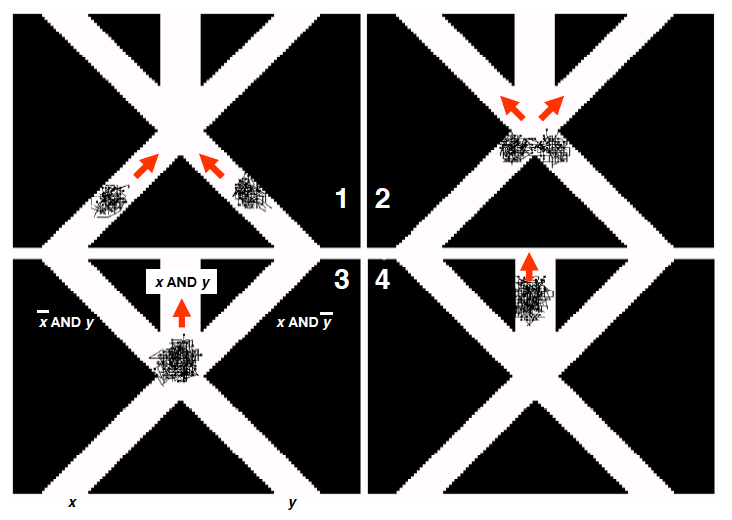}
\caption{A series of snap shots (1, 2, 3 and 4) in {\sc AND} gate of swarm balls. A swarm ball locate at $x$- and 
$y$-position consists of 40 agents, respectively. Each agent is represented by a square with its 5-steps-trajectories. Red arrows represent the direction of motion of a swarm ball.}
\label{fig03}
\end{figure}

Figure \ref{fig03} shows {\sc AND} gate of swarm balls. Given a pair of inputs $x$ and $y$ located below, the left, central and right corridor located above represents the output as {\sc NOT}$(x)$  {\sc AND} $y$, $x$ {\sc AND} $y$, and $x$ {\sc AND NOT} $(y)$, respectively. Fig. 3 shows a series of snapshots for the input $(x, y) = (1, 1)$. Two swarm balls initiated at input locations moves along the wall, leftward and rightward, respectively. Each swarm ball consists of forty agents. After the collision, a united swarm moves upward due to integration of velocities. 
It results in a united swarm moving to the central corridor located above. It shows that $x${\sc  AND} $y$ is 1. As well as the behaviour of {\sc OR}-gate, a united swarm ball consists of eighty agents and directions of transitions of agents are not analogous. A swarm ball continuously contains internal turbulence. Initial configurations of agents in these gates are randomly given in a designated area in the input position, where no external noise is coupled with the transition rule for the agent. The performance of these gates is 100 percent for 
{\sc OR} and, {\sc AND} (also {\sc NOT}) gates. Because we can implement {\sc OR}, {\sc AND}, and 
{\sc NOT} gates, we can calculate any propositional logic sentences.

Finally we discuss the robustness of the gates. As mentioned before, {\sc OR} gate acts very well against the external perturbation. Here we estimate the performance of {\sc AND} gate under a high rate of external noise. In the simulation, the external noise that is coupled with the operation of velocity matching is defined as a random value, $\xi$, selected with equal probability from $[-\lambda, \lambda]$. When the velocity of each agent in a two dimensional space, $v$, is projected in $x$- and $y$-plane and is denoted by $v_x$ and $v_y$, the external noise is coupled by $v_x+\xi$ and $v_y+\xi$ and then normalised to make the length of the velocity be 1. We estimate the performance within the range, $0.0 \leq \lambda \leq 0.2$, in comparing our model of $P=20$ with one with $P=1$. 
If $P=20$ multiple potential transitions play a role in mutual anticipation which can contribute to collective behaviour. If $P=1$, coherence of swarm results only from velocity matching.

\begin{figure}
\centering
\includegraphics[width=0.7\textwidth]{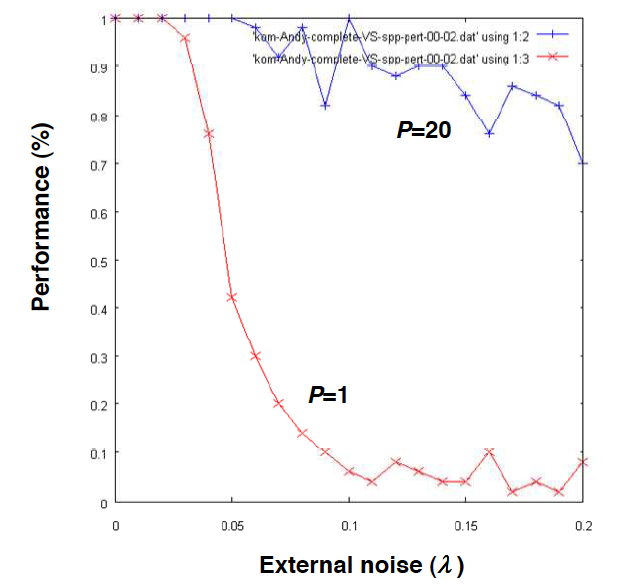}
\caption{Performance of {\sc AND} gate implemented by the swarm model. Performance is plotted against the strength of external perturbation. A swarm model with $P=20$ is compared with that with $P=1$. The model with $P=1$ can correspond to BOIDS.}
\label{fig04}
\end{figure}

Figure~\ref{fig04} shows performance of {\sc AND} gate implemented by our swarm model under a perturbed condition. The performance is defined by the success rate of {\sc AND}-gate. In each experiment we set 40 agents in the input position $x$ and $y$, respectively. If 80\% of agents (64 agents) in a united swarm reach the output exit of $x$ {\sc AND} $y$, we determine the result for the experiment as ÒsuccessÓ; otherwise, ÒfailureÓ. The performance is defined by the number of succeed experiment divided by the total number of experiments (100 experiments). Since the swarm model with $P=1$ corresponds to the well-known model BOIDS, the external perturbation directly influences the collective behaviour of swarms. Therefore, as the strength of perturbation increases, the performance rapidly decreases.

In contrast, the performance of our swarm model with $P=20$ does not decrease drastically as the strength of perturbation increases. A large possibility of transitions can increase the probability of presence of the high popular sites, which increases possibility of mutual anticipation. It results in the collective behaviour of swarming. External perturbation also increases the possibilities of transitions. In this sense the external perturbation cannot be distinguished from potential transitions, and can produce same contribution as increasing the number of potential transitions. The external perturbation plays a role in generating and keeping a swarm ball. The directed motion of a swarm along the wall, however, depends on the strength of velocity matching. That is the reason why the performance of {\sc AND} gate implemented by swarms with $P=20$ is slightly decreased. Note that the cohesive power does not weaken despite low performance.

\subsection{Experimental gate implemented by real soldier crabs}

We first implemented the logical gate of swarm collision by real soldier crabs, \emph{Mictyris guinotae}. The corridor is made of acrylic plastic plate, and the gradient to move was implemented by using the orange-coloured intimidation plate. The reason for having an intimidation plate is because in natural environment birds  
are main natural predators feeding on soldier crabs and there are usually now shadows in the lagoon except for the shadow of the birds.  This why the crabs are very sensitive to the shadows made by standing and moving object. Thus the intimidation plate can trigger soldier crabs to move away from the shadowed region. The floor of the experimental gate is made of cork to provide comfortable for crabs friction.

\begin{figure}
\centering
\includegraphics[width=0.7\textwidth]{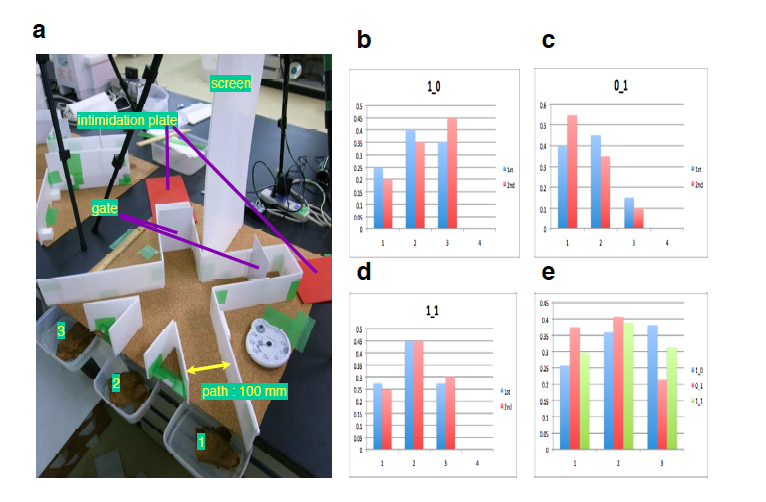}
\caption{a. Implementation of {\sc AND} gate by real soldier crabs. A or B represents input space for input $x$ or $y$, respectively. The symbol 1, 2 and 3 represent output for {\sc NOT}$(x)$ {\sc AND} $y$, $x$ {\sc AND} $y$, 
and $x$ {\sc AND} {\sc NOT}$(y)$, respectively. The experimental results of {\sc AND} gate implemented by real soldier crabs. b. Frequency distribution of output 1, 2 and 3 for input, (1, 0). Output 4 represents the rate of individuals not reaching the output in a limited time. c. Frequency distribution for input, (0, 1). d. Frequency distribution for input, (1, 1). e. Frequency distribution of output 1, 2 and 3 over 21 trials.}
\label{fig05}
\end{figure}

Figure \ref{fig05}a shows the {\sc AND} gate of swarm collision. First we close the gate and set a swarm of solider crabs for either of them or both inputs space surrounded by wall and gate. The crabs are left about two minutes to relax. After the relaxation, folded intimidation plate is extended and stands, and then crabs start moving along the corridor. Real-world crabs indeed behave similarly to simulated crabs.  If input for 
$(x, y)$ is $(1, 0)$ or $(0, 1)$, a swarm representing value of 1 is set by a swarm of forty crabs. If input for $(x, y)$ is (1, 1), each swarm representing value of 1 is set by twenty crabs.

Fig. 5b-e shows the two results of {\sc AND} gate experiment for a particular swarm consisting of forty individuals. First forty individuals are divided into two parts, twenty individuals each, and set for input space $(x, y) = (1, 1)$. In other cases where $(x, y) = (1, 0)$ or $(0, 1)$, forty individuals are set for input space, A or B, respectively. In any input cases, both intimidation plates are unfolded and stood vertically, soon after the gate is opened. The bar of Figure 5b-e shows the rate (\%) of individuals reaching the output 1, 2 or 3. Red bars or blue represent the result of the first and second experiment for the same population of individuals. Two experiments have the same tendency, such that for the input $(1, 0)$ and $(0, 1)$ most part of a swarm goes straight (i.e., A to 3 or B to 1 in Fig. 5b-d), and for the input (1, 1) individuals derived from input space A and B are united and go to the central part of output, called 2. These results are approximated by our model simulation.

Fig. 5e shows experimental results for 21 trials, where each trial was conducted for forty individuals. Each bar represents the rate over all trails. The frequency histogram shows that for input (1, 0) and (0, 1) individuals of a swarm go straight or go to the central corridor. It means that most of a swarm goes straight to some extent. It also shows that for input (1, 1) most of a swarm goes to the central corridor after the collision of swarms. If the output of either 0 or 1 is determined by the difference of the number of individuals reaching the output 1 and 3 such that the output is 0 if the difference exceeds 10\% of all individuals (i.e. four individuals); 1 otherwise. If the determinant is applied to 21 trails of {\sc AND} gate experiment, performance of the {\sc AND} gate for input (1, 0), (0, 1) or (1, 1) is 0.81, 0.76 or 0.52, respectively.

Although the experiment implemented by real soldier crabs is just a preliminary work, the performance of the {\sc AND} gate is not bad. Since the {\sc OR} gate can be more easily implemented, the swarm collision computation can be naturally constructed. The gradient of the corridor is tested by other devices such as illumination, physical gravitational stimulus produced by slope, and some combinations. The condition of a swarm in the experiment in a room is different from that in a natural lagoon. If the good enough condition for swarms is provided even for the experiment in a room, robustness of a swarm can be reconstructed in a room.

\section{Conclusion}

Soldier crabs live in the tropical lagoon wander as a huge swarm in the low tidal period, while they live under the sand in the high tidal period. Each individual sometimes wanders freely, and is aggregated into a huge swarm triggered by natural enemy or tidal movement dependent on the lagoon topography. Once a swarm is generated, individuals in the peripheral regions of a swarm dynamically exchange their places and the internal turbulence of the crab becomes responsible for the same swarm motion. Therefore, even if the boundary of a swarm is definitely sharp and smooth, the internal structure of the swarm is dynamical and random. The mutual anticipation can negotiate perturbation and the force to make an order.  In this paper we implement ballistic computing~\cite{adamatzky_2003} by using such a robust swarm. In the natural implementation of ballistic computation it is difficult to refer to robustness of computation. For example, ballistic computation implemented by Belousov-Zhabotinsky reactions is unstable without well-controlled condition \cite{delacycostello_2011}. Our model suggests that biological computation is more robust even under the perturbed environments.

\textbf{Ethical note} No specific license was required for this work. The duration of any single experiment was so short that each crab never reached dangerous level, that the crabs were kept in comfortable condition, and that after all experiments the crabs were released to their natural habitats. Furthermore, on visual inspection, no crabs appeared to have been injured or adversely affected by the experiments.

\end{document}